\documentclass[aps,prl,amsmath,amssymb,preprint]{revtex4-1}

\usepackage{graphicx}
\usepackage{dcolumn}
\usepackage{bm}
\usepackage[mathlines]{lineno}

\def\be{\begin{equation}}
\def\ee{\end{equation}}
\def\ba{\begin{aligned}}
\def\ea{\end{aligned}}

\begin{document}
\title[]{\bf{\large Anderson localization and extreme values in chaotic climate dynamics}}

\author{John T. Bruun}
\email[Corresponding author: ]{j.bruun@exeter.ac.uk}
\homepage[]{http://emps.exeter.ac.uk/mathematics/staff/jb1033}
 \affiliation{\small College of Engineering, Mathematics and Physical Sciences, University of Exeter, Exeter, UK. College of Life and Environmental Sciences, University of Exeter, Penryn Campus, Penryn, UK.}%
\author{Spiros N. Evangelou}%
\email[Second author: ]{sevagel@uoi.gr}
 \affiliation{\small Physics Department, University of Ioannina, Greece.}
\date{\today}
\begin{abstract}
This work is a generic advance in the study of delocalized (ergodic) to localized (non-ergodic) wave propagation phenomena in the presence of disorder. There is an urgent need to better understand the physics of extreme value process in the context of contemporary climate change. For earth system climate analysis General Circulation Model simulation sizes are rather small, 10 to 50 ensemble members due to computational burden while large ensembles are intrinsic to the study of Anderson localization. We merge universal transport approaches of Random Matrix Theory (RMT), described by the characteristic polynomial of random matrices, with the geometrical universal extremal types max stable limit law. A generic ensemble based random Hamiltonian approach allows a physical proof of state transition properties for extreme value processes. In this work Anderson localization is examined for the extreme tails of the related probability densities. We show that the Generalized Extreme Value (GEV) shape parameter $\xi$ is a diagnostic tool that accurately distinguishes localized from delocalized systems and this property should hold for all wave based transport phenomena.
\end{abstract}

\maketitle
\section{\label{Intro}Introduction}
We introduce a new way to view Anderson localization by focusing at the extremes of physical quantities rather than at the mean values and apply it to the climate. Our world’s climate is changing at an alarming rate 
\cite{Collins2012}. These changes can and appear to be causing multifaceted impacts which can alter societal resilience \cite{Cai2015}.
The future state of the climate can now be simulated based on highly sophisticated Global Circulation Models (GCM's) that involves a substantial computational effort to study extreme climate phenomena in the presence of disorder. The chaotic and turbulent phenomena of the Earth require a generic understanding to enable predictive capacity  \cite{Young2011,Tong1990,Feigenbaum1980}.
Universal extreme properties have been studied for random energy models and Burgers turbulence in settings where eigenmode interactions are specified \cite{Derrida1981,Bouchaud1997,Fyodorov2008}.
However the physical explanation of the type of extreme value processes that occurs remains incomplete and we need to better establish how these alter through system state transitions $ \kappa \rightarrow \kappa^{'}$, such as in tipping points of the climate system
\cite{Lenton2008}.
Anderson localization for the absence of wave propagation in solids was established in the late 1950’s (Anderson, 1958) \cite{Anderson1958}. The corresponding Anderson transition in the presence of disorder is explained via the scaling theory of Abrahams et al. (1979) \cite{Abrahams1979}. This ergodicity breaking transition is related to destructive wave interference and
for strong enough disorder implies the absence of wave transmission.It also provides us with a transport framework through which state transitions and extreme phenomena driven by the level of disorder can be studied \cite{Evers2008}.
On one side of the transition the delocalized random systems have correlated energy eigenvalues and ergodic wave flow called quantum chaotic which is understood via RMT \cite{Wigner1955,Evangelou1995}.
In the opposite limit of Anderson localization  the system eigenvalues are uncorrelated described by Poisson statistics. Recent work by Fyodorov et al. (2008, 2012, 2016) \cite{Fyodorov2008,Fyodorov2012,Fyodorov2016} has looked at extreme processes from the RMT perspective. The corresponding random matrices are described via their characteristic polynomial $D_N(E)$ which encapsulates all the system $N$ eigenvalues and energies $E$.

The Extreme Value Process (EVP) of any system can be characterised by the shape parameter $\xi$ that represents the extreme edge of the hysteresis characteristic \cite{Bruun1998,Gouldby2016,DeHaan1998,Bruun2017}
and it is widely used in design engineering such as flood prevention work to assess extrapolation and risk properties. The EVP distribution is justified from a linear renormalisation and extremal types or max stable limit law based on geometric universality
\cite{Leadbetter1983,DeHaan1984,Haan1994,Tawn1988}.
A result of this max stable limit law is that EVP systems, represented by the largest measurements $Z$ (such as the maximum of an ensemble), are described by the GEV distribution function $G(z;\xi)$ where the tail type: Weibull, Gumbel or Fréchet is distinguished by $\xi$. The type of distribution tells us the sensitivity of the tail process and provides a measure of how extreme variability changes in general for any dynamical system. In the context of system state changes $ \kappa \rightarrow \kappa^{'}$, (Young, 2011) \cite{Young2011} the extremes can alter. More erratic extremes under an altered scenario $\kappa^{'}$ such as in climatic change would correspond to a heavier tail $ \xi(\kappa^{'}) > \xi(\kappa)$. We extend the characteristic polynomial approach \cite{Fyodorov2008,Fyodorov2012,Fyodorov2016}
to generally establish the shape $\xi$ of extremes. We examine state transition properties for Anderson localization including superconductivity \cite{Lambert1993,Bruun1994a,Bruun1995}
and for a chaotic RMT diffusive system, such as a model of a black hole with added disorder \cite{Patel2019}

\section{\label{equivalence}Equivalence between climatic phenomena and electron disorder properties}
It is natural first to ask and discuss why classical wave systems for climate should have anything to do with properties related to the flow of electron waves in disordered media? The answer is quite simple: the wave properties of quantum electrons and the classical waves in climate follow the same physical laws so that techniques developed for wave propagation in the quantum world can be also explored in climate. Anderson localization for wave propagation in the presence of disorder also appears for ultrasound waves, microwaves, light, etc. \textit{The question of classical localization: a theory of white paint?} Anderson (1985) \cite{Anderson1985} explains the possibility of observing light localization in TiO2 samples. Anderson localization of classical waves turns out to be rather difficult to observe since nature applies some severe constraints, such as rather small cross-sectional scattering areas and absorption \cite{Lagendijk2009,Ludlam2005}.
These  properties of disordered  systems are thought to occur in many other settings: Anderson localization also occurs in many-body settings other than the real space of one particle. In strange metals Patel and Sachdeev  (2019) \cite{Patel2019} recently showed that significant amounts of disorder are present and electrical insulators governed by similar laws to chaotic metals exist. The Sachdeev-Ye-Kitaev model which connects Hamiltonian disorder models to the physics of the black holes offers a link of many-particle quantum entanglement to many-body localization
\cite{Altman2018}.
By analogy the physics and wave transport phenomena in the Earth system and the climate is regulated through disorder introduced through small scale wave activation processes and surface roughness. This can lead to phenomena such as convective self-aggregation and spatial clumping transport processes \cite{Wing2014}.
Recently there has been a resurgence of interest to assess wave properties of geophysical climate systems (Delplace, et al., 2017; Bruun et al., 2017; Skákala and Bruun, 2018) \cite{Delplace2017,Bruun2017,Skakala2018}. In these analyses universal and scale invariant properties help to identify the dynamics and the resulting wave processes. Equatorial waves in oceans which regulate climate have been shown to be driven by dynamics similar to the so-called topological insulators \cite{Delplace2017}
where unlike normal insulators a flow of states protected by symmetry occurs only on the surface of a sample. The bulk transport medium is insulating having localized states and delocalized waves unaffected by disorder travel around the edges of the system due to topology. For example in the 2D quantum Hall effect \cite{Sarma1997}
such edge states of electrons are protected by a strong magnetic field which breaks time-reversal symmetry and determine the highly accurate Hall conductance. In climate the equivalent role of the magnetic field is played by the Coriolis effect caused by planet’s rotation, and topological waves that are important for the dominant climatic processes on the Earth flow around the equator \cite{Delplace2017}. In particular wind induced oceanic Kelvin and Rossby waves travel along the Pacific equator, scatter and reflect at its edges and combine to create the Pacific El Niño Southern Oscillation (ENSO) resonance property \cite{Cane1985,Suarez1988,Tziperman1994,Munnich1991,Bruun2017,Skakala2018}.
Recently Bruun et al. (2017) \cite{Bruun2017} and Skákala and Bruun (2018) \cite{Skakala2018}
established that low frequency eigenmodes appear to be part of the ENSO process. They occur as a sub-harmonic resonance property and can alter systematically through a state transition parameter $\kappa$ that represents non-linear ocean-atmospheric coupling \cite{Feigenbaum1980,Tong1990,Hilborn2000,Young2001,Young2011,Bruun2017}.
A future warmer climate could alter the ENSO resonance through a change to $\kappa$ promting the question: \textit{Is the current instability of the ENSO modes an example of the hysteresis characteristic changing in the industrial period?} In other words is the ENSO extreme value process shape parameter changing? Here, we set up a novel framework that such questions can be addressed and possibly answered. With current GCM’s the ability to study large ensemble extremes systematically across changes of system state is not possible: a typical GCM ensemble has of the order of 10 to 50 members \cite{Collins2012}.
As such, a theoretical explanation of the EVP for a generic physical framework is prompted and the physics of electron wave transport provides such a framework.  Electrons and their absence (holes) have distinct wave dispersion properites, and the way in which they combine define the transport encountered in superconductivity. By analogy, in climate science, tropical ocean Kelvin and Rossby waves combine to produce the ENSO phenomena. Andreev reflection (Andreev, 1965) \cite{Andreev1965} in superconductivity is a wave interaction property where an electron entering a medium forms a Cooper pair which consists of an electron plus a hole which is retro-reflected as a hole outside
\cite{Bruun1994a,Bruun1995,Beenakker1997}.
The Andreev wave scattering and interaction properties for normal $s$-type superconductors can extended to topological current $p$-type disordered superconductors \cite{Qi2009}.
The Bogoliubov-de Gennes (BdG) Hamitonian (Bruun et al., 1994 and 1995)\cite{Bruun1994a,Bruun1995} represents this process. The full dynamical system structure is given  by appropriate $N \times N$ random matrices which discretize the available space to $N$ points. In the presence of a magnetic field a complex Hermitian GUE of RMT approximates a high-dimensional disordered systems, in this setting all to all interactions are included and distance plays no role. This GUE system exhibits highly correlated eigenvalues leading to universal statistical features \cite{Ludlam2005,Evangelou2000,Bruun1995,Wigner1955}
and for appropriate distributions for the maxima of the characteristic polynomial we identify the ergodic and non-ergodic state of the many-body system.

\section{\label{Hamiltonianequivalence}Hamiltonian eigenmode structure and extremes}
\begin{figure}[h!]
 \centering
   \includegraphics[scale=0.5]{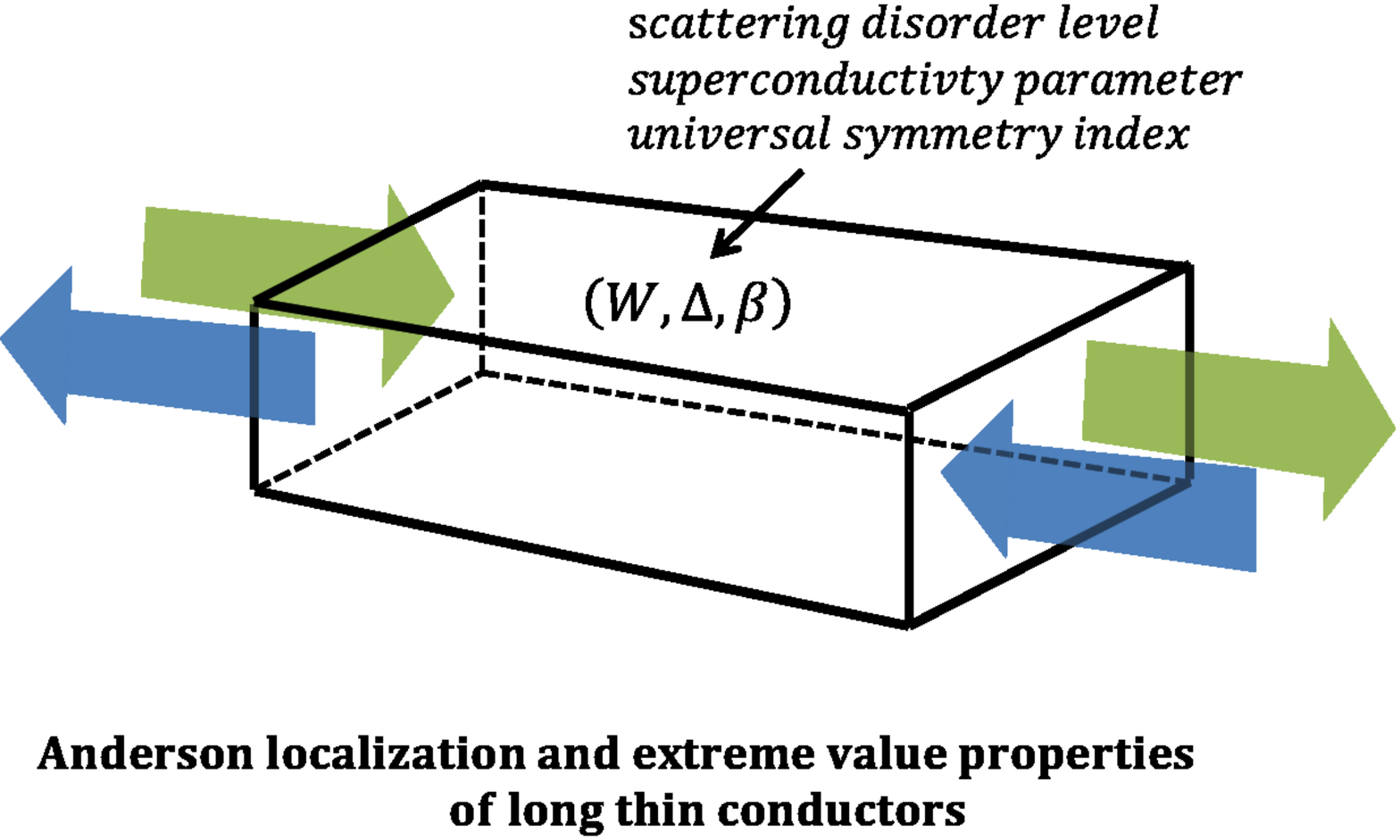}
    \caption{{\bf \footnotesize The finite quasi-1D setup.} {\footnotesize
    The transport flows in and out from the left to the right hand side of the conductor. The resistance and conductance properties which enable assessment of Anderson localization properties are studied by the characteristic polynomial of a similar closed system.
    }} \label{fig1}
\end{figure}

The Bogoliubov-de Gennes mean-field Hamiltonian equation reads
\begin{equation}\label{BdGeq}
\begin{aligned}
\left[ {\begin{array}{cc} H_{o}(W) & \Delta\\ -\Delta^{*} & -H_o^{*}(W)\\ \end{array}} \right] \underline{\psi} = \lambda \underline{\psi}
\end{aligned}
\end{equation}
$H_o (W)$ represents the electron-wave Hamiltonian in the presence of disorder specified by $W$, $-H_o^* (W)$ is the Cooper paired electron in the form of an Andreev reflected hole, $\Delta$ is the order parameter which represents the superconducting energy-gap that opens in the transport band structure when the temperature $T$ becomes lower than $T_c$ and the material becomes superconducting. The coupling matrix between electrons and holes for normal $s$-type and for topological $p$-type superconductors is
\be\label{Couplingeq}
\ba
\Delta =
\left( {\begin{array}{ccc} \Delta_{1} & & 0 \\
                              &  \Delta_{2} & \\
                             0 &  & \ddots \\
                           \end{array}} \right),
\Delta =
\left( {\begin{array}{ccc} 0 & -\Delta_{1} & \vdots \\
                            \Delta_{1}^{*} & 0 & -\Delta_{2} \\
                             \vdots & \Delta_{2}^{*} & \ddots \\
                           \end{array}} \right)
\ea
\ee
respectively. In the case of $\Delta=0$ the system is not superconducting and the electron Hamiltonian $H_o(W)$ can exhibit Anderson localization properties depending on the disorder $W$. The studied system is shown in Fig.~\ref{fig1}.
In the delocalized regime the $N \times N$ Hermitian matrix $H_0$ (the Hamiltonian) has a probability density function for Gaussian random matrix ensembles of RMT
\be\label{GUEeq}
\ba
P(H) \propto exp\{-(N \beta/4) Tr(H^{2})\}
\ea
\ee
and takes various forms depending on the universality parameter $\beta =1,2,4$. The  $\beta=2$ case studied here corresponds to the unitary GUE limit where time-reversal symmetry is broken. The characteristic polynomial of the matrix $H$ at an energy $E$ is obtained from the determinant
\be\label{DNEeq}
\ba
D_{N}(E)= det( EI - H) = \prod_{j=1}^{N} (E - E_{j}),
\ea
\ee
where $I$ is the $N \times N$ unit matrix. $D_N(E)$ is the basic quantity of interest encoding all eigenvalues $\{E_j  ; j=1,...,N\}$ and the roots of $H$ are obtained from matrix diagonalization. The transport quantity of interest is the resistance of the system \cite{Gasparian1988,PhysRevLett.78.2803}
\be\label{RNeq}
\ba
R_N(E)=|D_N (E)|^{2} -1.
\ea
\ee
Following the definition by Fyodorov and Simm (2016) \cite{Fyodorov2016} $D_N(E)$ is expressed as
\be\label{fDNeq}
\ba
f\{D_{N}(E)\}= |D_{N}(E)| exp \{-<log|D_{N}(E)|>\}
\ea
\ee
where $<...>$ is average over the ensemble. The extreme value processes (EVP) is an intrinsic property of the characteristic polynomial and the maximum embeds the characteristic of the $N$ eigenvalues for a given ensemble $i$ as
\be\label{MiNeq1}
\ba
M_{i,N}=max_{E \in [-2,2]} \{ 2 log |D_{i,N}(E)| - 2 <log |D_{i,N}(E)|> \}.
\ea
\ee
We create $n$ ensembles of this process so the set of ensemble maxima are
\be\label{MiNeq2}
\ba
M_{n}=\{M_{1,N},M_{2,N},...,M_{n,N}\}.
\ea
\ee
A linear renormalisation extremal types theorem (Leadbetter, 1983) \cite{Leadbetter1983} converts these maxima  (\ref{MiNeq2}) into a more useful representation by scaling the variable as $M_n^{*}=(M_{n} - b_{n}) / a_{n}$.  The max stable limit theorem (similar to the central limit theorem) gives
\be\label{MaxStableeq}
\ba
Pr\{(M_{n}-b_{n}) / a_{n} \leq z\} \rightarrow G(z),  n \rightarrow \infty,
\ea
\ee
where the selection of $\{a_{n}\}$ and $\{b_{n} \}$   results in the max stable limiting distribution $G(z)$ for limit of $M_n^{*}$. This distribution is generic and does not depend on any individual generating distribution function such as the disorder process of strength $W$. A distribution is called max-stable if for every $n=2,3,...$ the constants $a_{n}>0$ and $b_{n}$ exist such that
\be\label{MaxStable2eq}
\ba
G^{n} (a_{n} z+ b_{n} )=G(z),   n \rightarrow \infty,
\ea
\ee
so the max stability property is satisfied by distributions for which the operation of taking sample maxima leads to an identical distribution, apart from a change of scale and location. The apparent difficulty that normalising constants will be unknown in practice is easily resolved due to the limit in (\ref{MaxStable2eq}) as
\be\label{MaxStable3eq}
\ba
Pr\{ (M_{n}-b_{n}) / a_{n}  \leq z\} \approx G(z)
\ea
\ee
for large enough $n$, so equivalently
\be\label{MaxStable4eq}
\ba
Pr\{  M_{n} \leq z\} \approx G\{(z-b_{n})/a_{n} \}=G^{*}(z),
\ea
\ee
where $G^{*}$ is another member of the same GEV family. This extremal types theorem enables approximation of the distribution of $M_{n}^{*}$ by a member of the GEV family for large $n$, and so the distribution of $M_{n}$ itself can be approximated by a different member of the same family. This is a useful property for the estimation stage. As the parameters of the distributions $G$ and $G^{*}$ have to be estimated, it is irrelevant in practice that the location and scale parameters of the distributions will be different. The type of GEV distribution, defined by its shape $\xi$, will be the same. Due to this the properties of a max stable process are estimated using the GEV parameterisation
\be\label{GEVeq}
\ba
G(z) = exp \{-\left[1 + \xi (z - \mu)/\sigma \right]^{-1/\xi}\} = GEV(z; \mu, \sigma,\xi)
\ea
\ee
using likelihood or rank based inference \cite{Coles2001,Tawn1988,DeHaan1984,Haan1994,Bruun1998,DeHaan1998}.
As the shape is the invariant term we refer to (\ref{GEVeq}) as $G(z; \xi)$.  The location parameter has $\mu \in \mathbb{R}$, the scale parameter $\sigma >0$ and shape parameter $\xi \in \mathbb{R}$ and $
1 + \xi (z - \mu )/ \sigma  > 0 $. For $\xi=0$, the distribution simplifies to a Gumbel or Type I distribution:
\be\label{Gumbeleq}
\ba
G(z)= exp \{-exp \left[-(z - \mu)/\sigma\right]\}
\ea
\ee
which is unbounded. For $\xi >0$, it is known as a Fréchet or Type II distribution, with a bounded lower tail at $z=\mu-\sigma /\xi $ and infinite upper end point. A Fréchet distribution has a heavy upper tail. For $\xi <0$, it is known as the Weibull or Type III distribution, with a bounded upper tail at $z=\mu -\sigma/\xi$ and infinite lower end point. The GEV type properties of these maxima are evaluated below. To estimate the parameters $\mu$, $\sigma$, $\xi$ we use the maximum likelihood estimation approach. The likelihood $L$ is the joint probability of all the ensemble members occurring with the given probability is:
\be\label{Likelihoodeq}
\ba
L(\mu,\sigma,\xi)=\prod_{i=1}^{n} Pr\{Z = z_{i};\mu,\sigma,\xi\}.
\ea
\ee
Given the GEV parametrisation of Eq. (\ref{GEVeq}) and the ensemble of maximum data, the EVP is estimated by optimising $l = logL$ the log-likelihood function of (\ref{Likelihoodeq}) as
\be\label{loglikelihoodeq}
\ba
l(\mu,\sigma,\xi)= -n log\sigma -(1+1/\xi)\sum_{i=1}^{n}log\left[1+\xi(z_{i} -\mu )/\sigma \right]  \\- \sum_{i=1}^{n}log\left[1+\xi(z_{i}-\mu)/\sigma\right]^{-1/\xi} .
\ea
\ee
The estimates are obtained numerically as analytical maximisation is not possible. The standard errors, confidence intervals and profile log-likelihood of the shape $l(\xi)$ (which allows specific testing of the shape parameter sign) are obtained from (\ref{loglikelihoodeq}) using standard likelihood theory \cite{Coles2001}.
The principle of maximum likelihood estimation for a suitable ensemble is to adopt the set of parameters with the greatest likelihood, since of all the range of parameter combinations, this is the one which assigns highest probability to the observed ensemble. This likelihood estimation approach is asymptotically fully efficient, i.e. no other estimation approaches can have a smaller estimation variance \cite{Pawitan2013,Coles2001},
so we can be confident that the estimated values (and the standard errors) of $\mu$, $\sigma$, $\xi$ are highly accurate.  In practice ensemble sizes $n \sim 1000$ obtainable with these random Hamiltonians provide the effective limit to derive the universal properties.

\section{\label{wavescattering}Wave scattering and state transitions}
\subsection{\label{localization}Localization}
\begin{figure}[h!]
 \centering
   \includegraphics[scale=0.6]{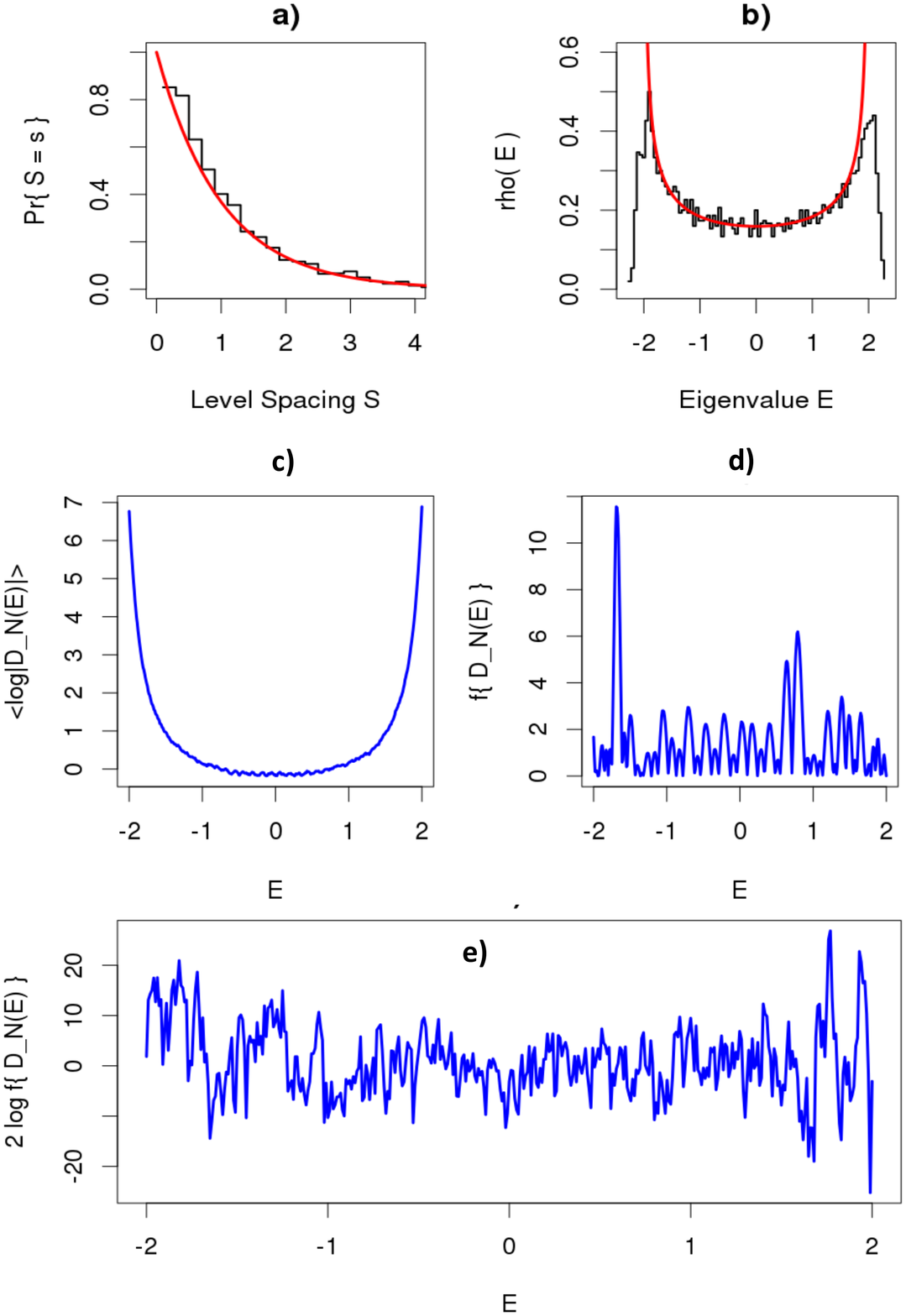}
    \caption{{\bf \footnotesize 1D Anderson localization} {\footnotesize
    $W=1$. a) $P(s)$ the level spacing between consecutive eigenvalues (red line is the Poisson distribution $exp\{-S\}$ for localized states). b) Eigenvalue distribution for $1D$ ballistic states, $N=3000$. c) The ensemble $<log|D_{N}(E)|>$ property and characteristic polynomial $f\{D_{N}(E)\}$ for two $1D$ chains: d) $N=50$ for ballistic and e) $N=3000$ for localized systems.}}
    \label{fig2}
\end{figure}

\begin{figure}[h!]
 \centering
   \includegraphics[scale=0.6]{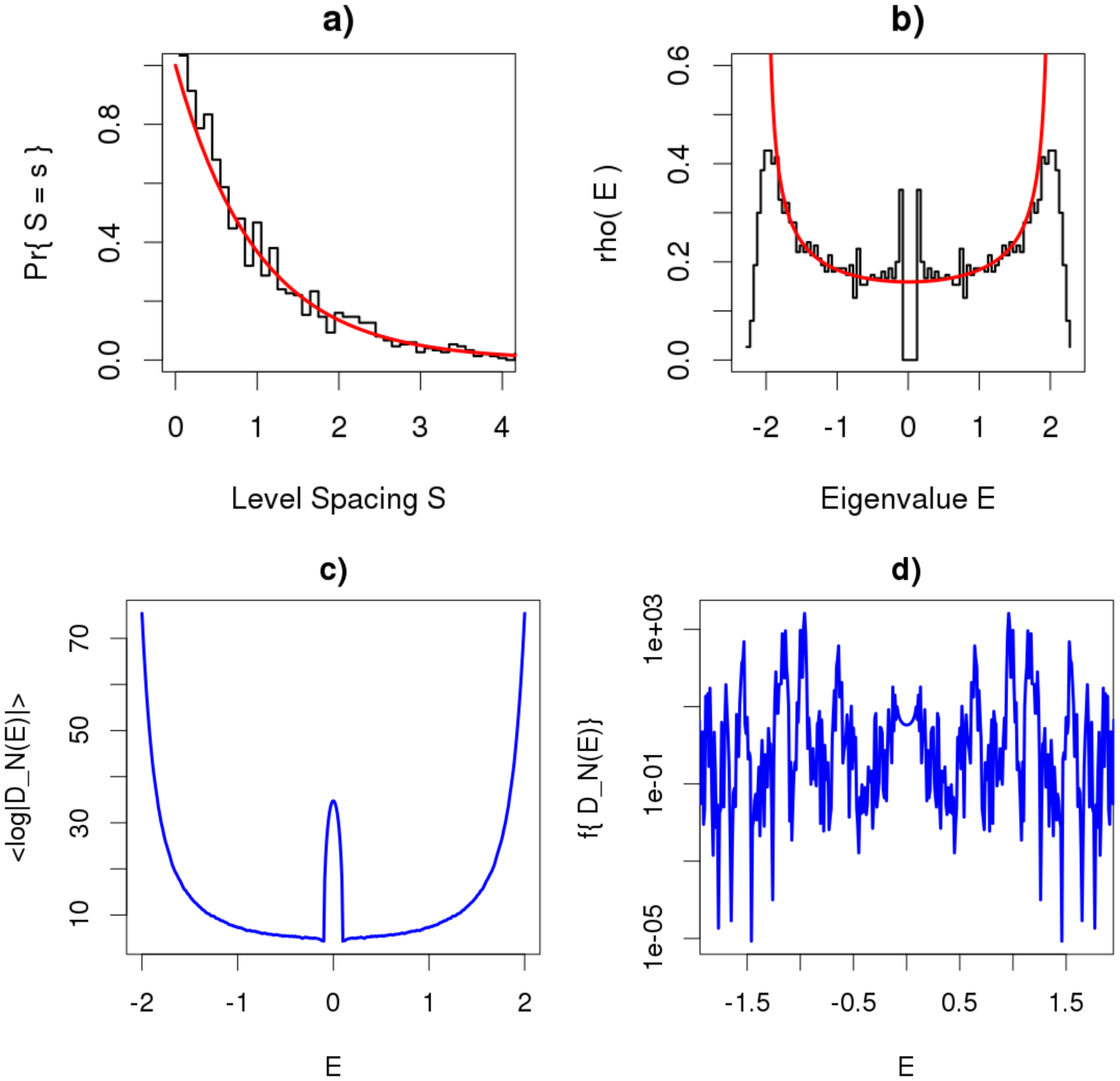}
    \caption{{\bf \footnotesize : Superconductivity} {\footnotesize
    with $\Delta =0.1$. a), b) the level spacing and eigenvalue Poisson distribution for the 1D localized limit with $N=1500$. c) The ensemble $<log|D_{N}(E)|>$ property and d) characteristic polynomial $f\{D_{N}(E)\}$ for $N=500$. The superconducting energy gap near $E=0$ is visible.}}
    \label{fig3}
\end{figure}

The system is described by $H_o(W)$ consists of $N$ sites arranged in a chain with random site potential $V_{j},j=1,2,...,N$ and $t_{j,j \pm 1}$  hoppings between the nearest-neighbour sites $j,j \pm 1$. The $1D$ Anderson model \cite{Anderson1958}
 is defined by a tridiagonal random matrix with diagonal matrix elements $V_{j}, j=1,2,...,N$ and off-diagonal matrix elements $t_{j-1,j}$ above and below the main diagonal. We take all the hoppings $t_{j-1,j}=1$ which defines the energy scale and the site potentials $V_{j} \in [-W/2,W/2]$ are independent random variables identically distributed with a uniform probability distribution of width $W$ which represents the strength of the disorder. The eigenvalues of the random system described by $H$ are $E_{j},j=1,2,...,N$ and the characteristic polynomial is evaluated via (\ref{DNEeq}). In Fig.~\ref{fig2} a) for a system of size $N=3000$ the eigenvalue level spacing $P(s)$ distribution with $S=(E_{j}-E_{j-1})/ <S>$ is shown.
The $W=1$ setting corresponds to a Poisson configuration with $Pr\{S=s\} = exp(-s)$. In b) the eigenvalue density $\rho(E)$ shows the system is close to the ballistic limit $\rho(E)=1/\pi \sqrt{1/(4-E^{2})}$ with singularities at $E = \pm 2$. Fig.~\ref{fig2} c) shows ensemble property $<log |D_{N}(E)|>$ for $n=3000$ verses energy and two single ensemble members of the characteristic polynomial are shown for d) $N=50$ and d) $N=3000$. Fig.~\ref{fig3} shows the superconductivity $s$-type case for $N=500$ in the $1D$ limit (with uniform disorder, $W=1$ and $\Delta=0.1$).
The level spacing distribution a) shows the Poisson limit. The superconducting energy gap is clearly evident in Fig.~\ref{fig3} b) the density of eigenvalues $\rho(E)$, c) the ensemble property $<log |D_{N}(E)|>$ and d) the characteristic polynomial. It is interesting to note that the density of states $\rho(E)$  shows both the near ballistic limit Poisson characteristic and the superconducting energy gap at $E =0$. Also note how the additional symmetry imposed by superconductivity with a real order parameter has reduced the complexity of the characteristic polynomial to be symmetric around $E=0$.

\subsection{\label{Delocalization}Delocalization}
\begin{figure}[h!]
 \centering
   \includegraphics[scale=0.6]{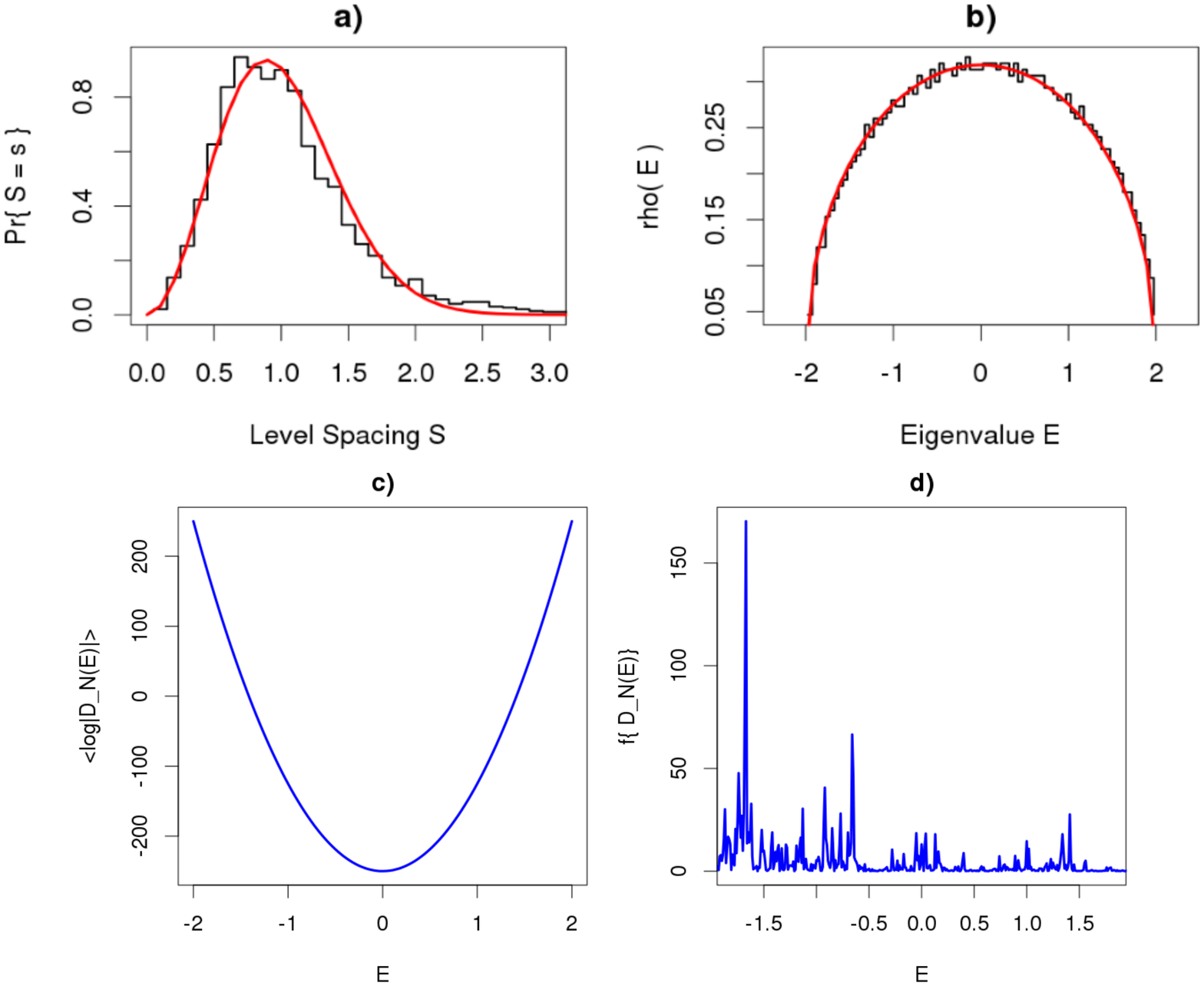}
    \caption{{\bf \footnotesize: Random matrix theory ergodic limit.} {\footnotesize
    The a) GUE level spacing distribution $P(S)=\frac{32 S^{2}}{\pi^{2}}exp(-4 S^{2} /\pi)$ and b) the Wigner semicircle eigenvalue density with $N=3000$. c) the ensemble $<log|D_{N}(E)|>$ property and d) characteristic polynomial $f\{D_{N}(E)\}$.}}
    \label{fig4}
\end{figure}

For a general GUE system $N \times N$ random Hermitian matrix $H_{0}=H_{0}^{\dagger}$ ($\dagger$ is the conjugate transpose) the entries are sampled from a Gaussian process, $N(\mu,\sigma^{2})$, $\mu$ the mean and $\sigma^{2}$ the variance. The diagonal matrix elements are
\be\label{GUEH1eq}
\ba
H_{i,i} \sim N(0,1), i = 1,...,N
\ea
\ee
and the off-diagonal matrix elements for $i>j$ have
\be\label{GUEH2eq}
\ba
H_{i,j} = x_{i,j} + \underline{i}y_{i,j};    x_{i,j},y_{i,j}  \sim N(0,1/2).
\ea
\ee
This sampling approach ensures Hermitian symmetry. To find the eigenvalues $E_{j}, j=1,...,N$ within $\pm 2$ the random matrix entries are scaled by $\sqrt N$. In Fig.\ref{fig4} a) we show the GUE level distribution universal characteristic for $N=3000$.
Note how $Pr\{S=0\}=0$, i.e. there is eigenvalue repulsion in this system and its properties are very distinct to that of the Poisson characteristic in the near ballistic limit. In b) the limiting mean density of the GUE eigenvalues is shown which is given by the Wigner semicircle law $\rho (E)=1/2\pi \sqrt{4-E^{2}}$ supported on the interval $E \in [-2,2]$. The ensemble property and the characteristic polynomial of one ensemble are shown in c) and d).

\subsection{\label{EVP}Extreme value process $\xi$ by $\kappa$ type}
\begin{figure}[h!]
  \centering
   \includegraphics[scale=0.6]{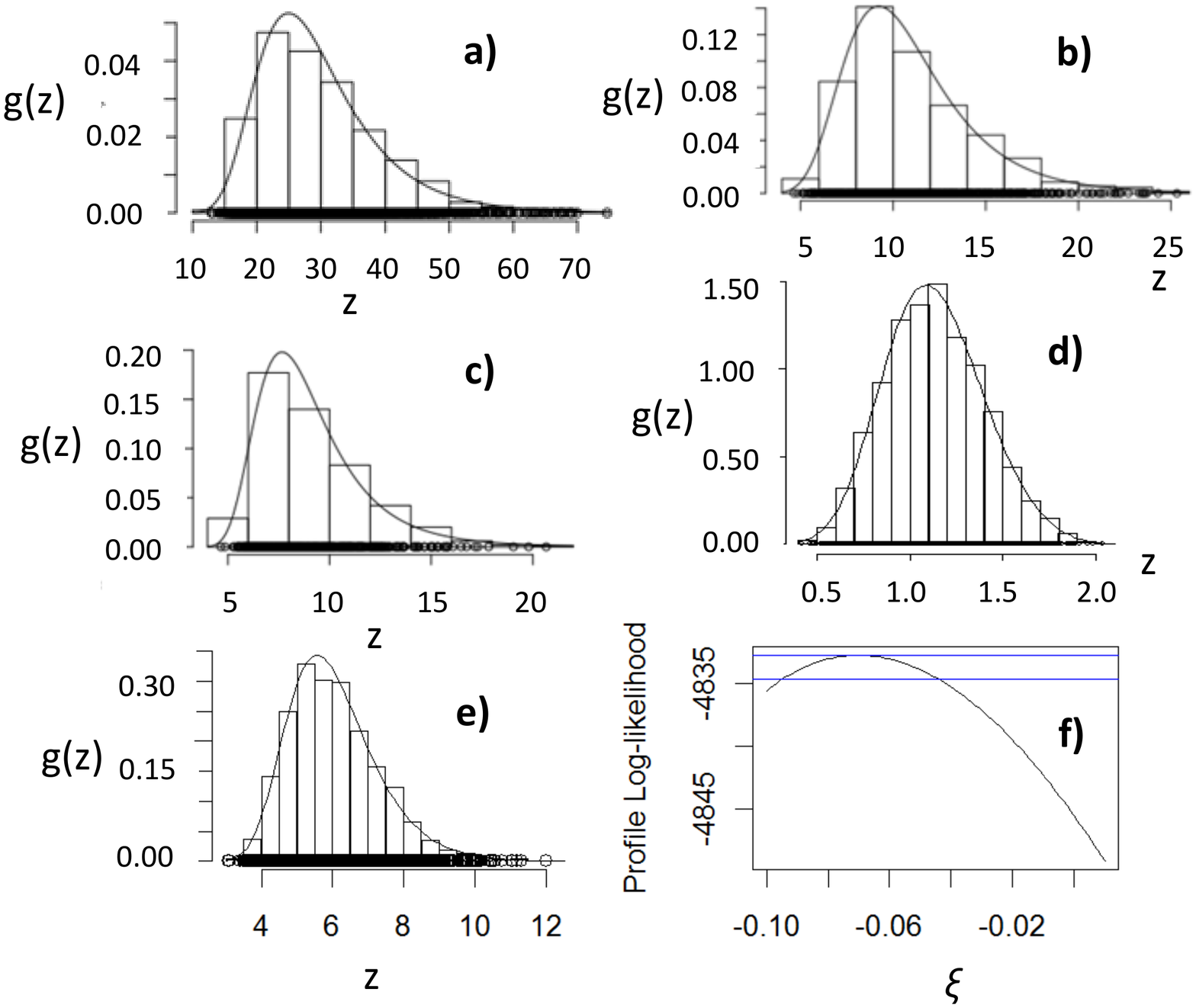}
    \caption{{\bf \footnotesize GEV distributions of ensemble maxima.} {\footnotesize
    a) to e) histogram and estimated GEV density (black line). $1D$ systems: a) $W=1$ electron system, $N = 3000$ and $n = 3000$. b) superconductivity with $\Delta=\Delta_{0}$ real, $W = 1$, $N=300$ and $n=1000$. c) superconductivity with $\Delta = \Delta_{0}e^{\underline{i}\theta}$ random phases $1D$ limit, $W = 0.1$, $\Delta_{0} = 0.3$, $N=300$ and $n=500$. d) $1D$ Ballistic electron limit $W = 0.01$, $N=50$ and $n=1000$. e) GUE system with $N=500$ and $n=3000$. f) For case e) the profile log-likelihood $l(\xi)$ (black line) of $\xi$, the 95\% confidence interval range is given by the drop of the lower blue line (from the maximum) indicating $\xi < 0$.}}
    \label{fig5}
\end{figure}

We have assessed the shape parameter across a range of state transitions. Our study is free from the assumption that the GEV shape parameter $\xi=0$ with $G(z;  \xi=0)$ known as the Type I or Gumbel extreme value limit
\cite{Fyodorov2008,Fyodorov2012,Fyodorov2016}.
In Fig.~\ref{fig5} we show the estimated extreme value distribution function $G(z;\xi )$ (\ref{GEVeq}) and the associated $\xi$ is given in Table~\ref{table:extremes}.
\begin{table}[ht]
\caption{The EVP shape parameter $\xi$ by state transition and process type.} 
\centering 
\begin{tabular}{c c c c c c c} 
\hline\hline 
 & $N$ & Ensemble & Disorder & Order     & GEV $\xi$  & GEV type \\ [0.5ex] 
 &     & size $n$ &          & parameter &($95\% CI$) &          \\ [0.5ex] 
\bf{$1D$ strong disorder}& 500 & 3000 & $W=1$ & na &0.077&Fréchet\\ 
                         &     &      &       &    &(0.046, 0.107)&\\ 
\bf{$1D$ weak disorder} & 500 & 1000 & $W=0.01$ & na &-0.211&Weibull\\
                        &     &      &          &    &(-0.233, -0.189)&\\
\bf{1D superconducting} & 300 & 1000 & $W=1$ & $\Delta_0=0.1$ &0.043&(Fréchet)\\
                        &     &      &       &                &(-0.007, 0.093)&\\
\bf{} &300 & 500 & $W=1$ &$\Delta_0=0.3$           &0.103&Fréchet\\
      &    &     &       &$\theta \sim U(-\pi, \pi)$ &(0.008, 0.184)&\\
\bf{delocalized RMT limit} &500 & 3000 & GUE & na &-0.070& Weibull\\
                           &    &      &     &    &(-0.095, -0.045)&\\[1ex] 
\hline 
\end{tabular}
\label{table:extremes} 
\end{table}
In the $1D$ setting for the electron Hamiltonian with $W=1$ (Fig.~\ref{fig5} a) the shape parameter $\xi>0$ indicating a Fréchet tail and that the system has a finite lower bound at
$z = \mu -\sigma / \xi$. For the same system but with $s$-type superconductivity (Fig.~\ref{fig5} b: real order parameter) the shape parameter $\xi > 0$ which could indicate a Fréchet type, however the $95\%$ confidence inverval contains $0$ so a Gumbel tail characteristic may be present. When the superconducting order parameter becomes complex (Fig.~\ref{fig5} c) then $\xi > 0$ indicating a Fréchet type tail, so localization has occurred. This heavy tail result is consistent with that previously established for the $1D$ localized regime \cite{Somoza2007,Lambert1993,Bruun1994b}.
When the disorder level is reduced to the ballistic limit in the $1D$ Anderson transport setting (Fig.~\ref{fig5} d) then $\xi$ much less than zero indicating a Weibull type. Then the upper tail becomes lighter than for a Gumbel situation and the characteristic polynomial has an upper bound at
$z = \mu - \sigma / \xi$.
 This appears consistent with random energy model discussion of Bouchaud and Mezard (1997) \cite{Bouchaud1997} for systems with correlated eigenvalues.  For the GUE delocalized electron case (Fig.~\ref{fig5} e) $\xi <0$ indicating a Weibull type. This $D_{N}(E)$ property is an interesting example of a log-correlated Gaussian random field (Fyodorov et al., 2008) \cite{Fyodorov2008} and a partially proved conjecture for $max\{<log|D_{N}(E)|>\}$, in GUE can be found in Fyodorov et al. (2016) \cite{Fyodorov2016}. The delocalization property and its correlated eigenvalues appear to support the Weibull type for the GUE setting. We test this property with the profile log-likelihood $l(\xi)$  obtained from Eq. (\ref{loglikelihoodeq}) in Fig.~\ref{fig5} f) (the black line). The $95\%$ confidence interval is given by the width of the log-likelihood surface where it intersects the lower blue line. This confirms the shape parameter is negative (GEV diagnostics in supplementary: Fig. A1). Our results show that the extreme shape varies with the system state property $\kappa$. In particular for the transition from localized to de-localized, the max stable limit property (\ref{MaxStableeq},\ref{GEVeq}) implies that the mobility edge corresponds to the Gumbel type with $\xi \sim 0$. This work establishes the $\xi(\kappa)$ structure using the full GEV representation and the EVP for a range of universal symmetry breaking changes $\kappa \rightarrow \kappa^{'}$, known to exist in quantum transport problems. It follows generically from our analysis and the proof above that in extreme value process studies it is essential to include an assessment of $\xi$ the shape parameter. In climate systems we propose that changes in system state in the extremes of future climate will be measurable in the magnitude and sign of the shape parameter. For example a transition to a heavy tailed process could indicate a form of localization (Ludlam et al., 2005) \cite{Ludlam2005} in the associated climatic phenomena.

\section{\label{conclusions}Conclusions}
The interesting possibility arises that some of the phenomena observed due to wave propagation in climate could be a consequence of disorder. We link Anderson localization, a phenomenon at the heart of quantum physics, to the analysis of the climate. Anderson localization wave properties in disordered media, although difficult to observe for classical waves, arises from multiple wave scattering through a disordered medium. We define a new way to view the Anderson transition by focusing on the extremes of physical quantities rather than at the mean values and their fluctuations. Our approach is developed in low dimensional localized and infinite dimensional delocalized systems to assess a wide range of extreme physical processes. We have shown that $\xi>0$ occurs when the system is localized and $\xi<0$ when the system is delocalized and so via the max stable limit law the mobility edge at the Anderson transition should occur when $\xi \sim 0$. Our results can be extended to include non-ergodic multifractal states known at criticality between localized and delocalized regimes. More generally we have established that $\xi(\kappa)$ can change when the dynamical system fundamentally changes its physical structure $\kappa \rightarrow \kappa^{'}$ and that this is a universal result. As a consequence we can assess the extreme shape parameter $\xi$ of other systems, such as in the earth’s climatic system to help better characterise extreme risk scenarios. In conclusion, we have shown via the max stable law how to study Anderson localization encountered for wave phenomena in many random settings including climate.

\section*{Acknowledgements}
The authors gratefully acknowledge the UK Research Councils funded Models2Decisions grant (M2DPP035: EP/P01677411), ReCICLE (NE/M00412011) and Newton Funded China Services Partnership (CSSP grant: DN321519) which helped fund this research. The authors also acknowledge useful discussion with Katy L. Sheen on the Earth system transport context. The data used in this manuscript is simulated from the analytical formula stated in the manuscript using R and the extreme value likelihood inference with the R library \textit{ismev}. JTB and SNE designed the analytical and theoretical basis for this study. JTB set the wider geophysical context of the work performed the analyses and wrote the initial manuscript. Both the authors contributed to the analysis discussion and the completed manuscript. The authors declare no competing interests.

\bibliography{AndersonLocalization}{}
\bibliographystyle{apsrev4-1}
\end{document}


\title[]{\bf{\large Supplemental material:\\Anderson localization and extreme values in chaotic climate dynamics}}

\author{John T. Bruun}
 \affiliation{\small College of Engineering, Mathematics and Physical Sciences, University of Exeter, Exeter, UK. College of Life and Environmental Sciences, University of Exeter, Penryn Campus, Penryn, UK.}%
\author{Spiros N. Evangelou}%
 \affiliation{\small Physics Department, University of Ioannina, Greece.}
\date{\today}
\maketitle

\section*{\textbf{1D system characteristic polynomial via recursion}}
For the $1D$ system assessed in this work the characteristic polynomial can also be derived via the recursion relation
\be
\ba
D_{j} (E)=(E- V_{j} ) D_{j} (E)-t_{j-1,j}^2 D_{j-2} (E),     j=1,2,...,N.\\
D_{-1} (E)=0,D_{0} (E)=1.
\ea
\tag{A.1}\label{DNEeqRecusive}
\ee
The corresponding determinants are
\be
\ba
D_{1} (E)=E-V_{1}=(E- V_{1} ) D_{0} (E) \\
D_{2} (E)=
\left|
\begin{array}{cc}
E - V_{1} & -t_{12} \\
-t_{12}   &  E - V_{2}
 \end{array}
\right|
= (E- V_{2} ) D_{1} (E)-t_{1,2}^2 D_{0} (E)
\\
D_{3} (E)=
\left|
\begin{array}{ccc}
E-V_{1}  & -t_{1,2} & 0\\
-t_{1,2} &  E-V_{2} &-t_{2,3}\\
0        & -t_{2,3} & E-V_{3}
 \end{array}
\right|
=(E- V_{3} ) D_{2} (E)-t_{2,3}^2 D_{1} (E),...
\ea
\tag{A.2}\label{DNEeqRecusiveDet}
\ee

This gives the same results as for (4).
\renewcommand{\figurename}{Figure A}
\begin{figure}[h!]
  \centering
   \includegraphics[scale=0.65]{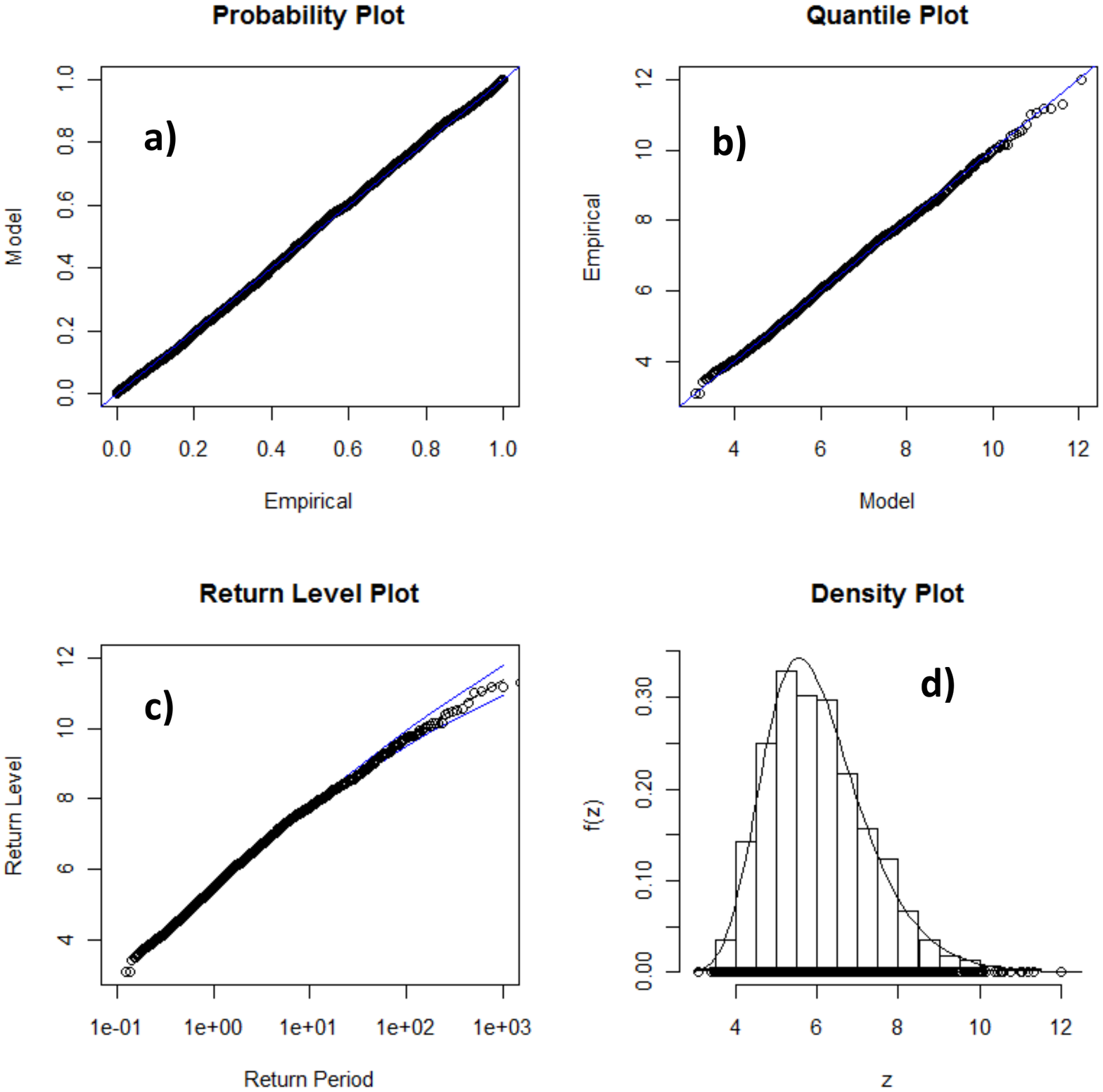}
    \caption{{\bf \footnotesize Diagnostics for GUE ensemble shape parameter (for Fig 5 e and f).} {\footnotesize
     The a) probability-probability and b) quantile-quantile plots here indicate that the GEV parametric model represent the raw ensemble of maxima well. c) The return level and the 95\% confidence interval range show the small negative shape parameter as a curvature in that graph. d) The ensemble histogram and fitted GEV (solid line).}}
     \label{figA1}
\end{figure}

\begin{figure}[h!]
  \centering
   \includegraphics[scale=0.65]{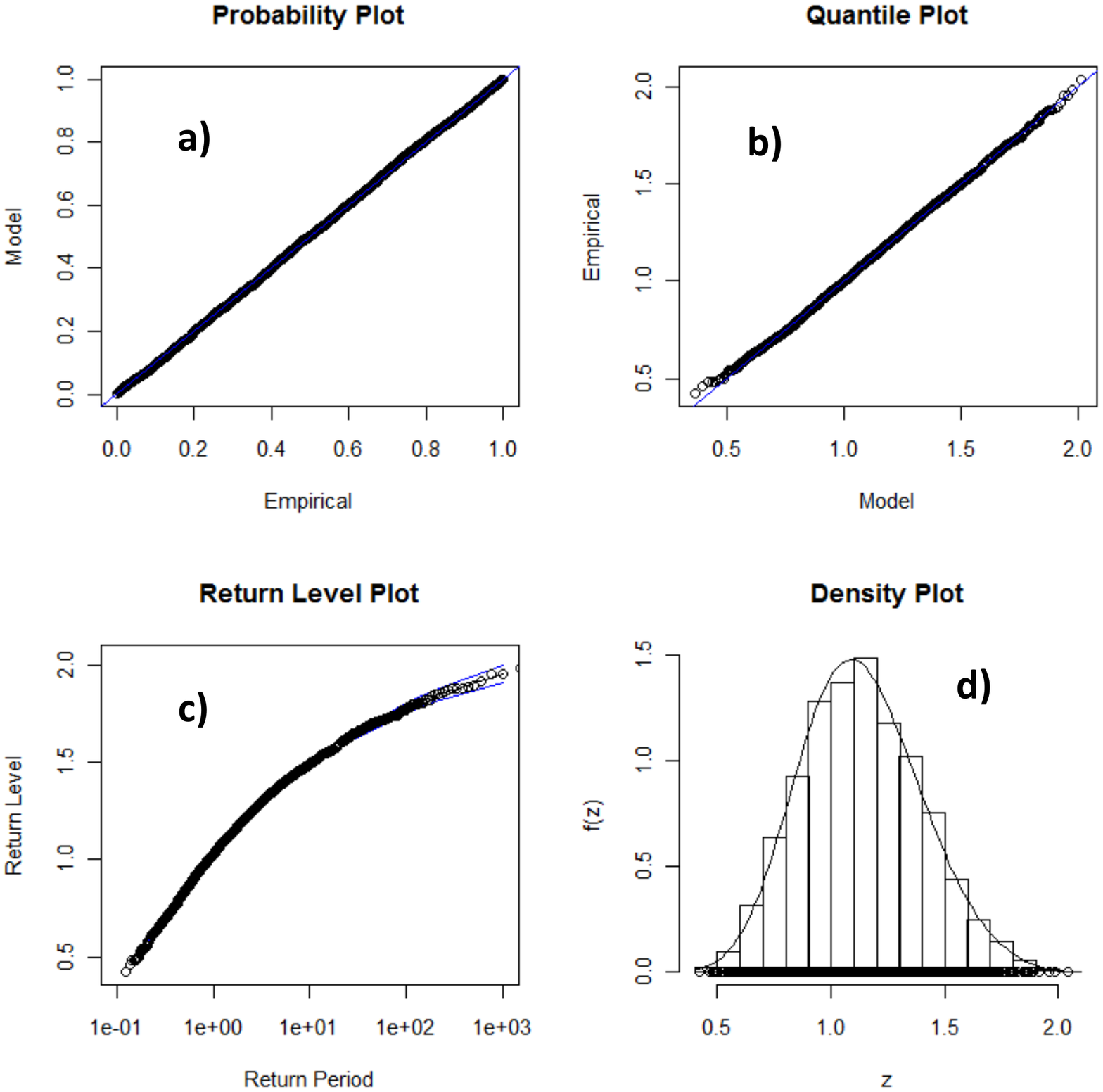}
    \caption{{\bf \footnotesize Diagnostics for $1D$ ballistic case. $W = 0.01$, $n = 1000$, $N = 500$.} {\footnotesize
     The a) probability-probability and b) quantile-quantile plots here indicate that the GEV parametric model represent the raw ensemble of maxima well. c) The negative shape parameter is evident in the curvature of the return level. d) The ensemble histogram and fitted GEV (solid line).}}
    \label{figA2}
\end{figure}

\begin{figure}[h!]
  \centering
   \includegraphics[scale=0.55]{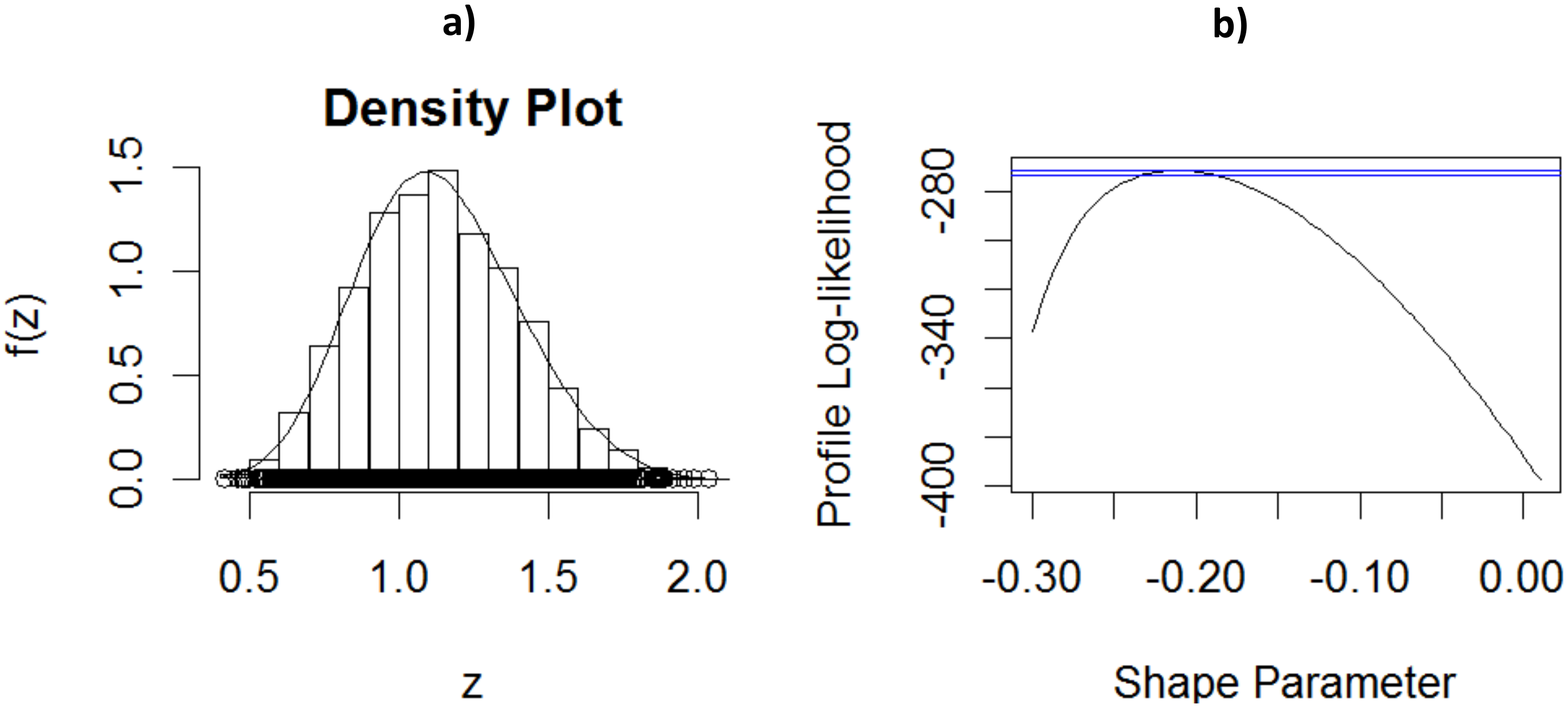}
    \caption{{\bf \footnotesize Diagnostics for $1D$ ballistic case. $W = 0.01$, $n = 1000$, $N = 500.$} {\footnotesize a) The histogram and fitted GEV (solid line). b) This shows the profile log-likelihood of the shape parameter and that this ensemble is a Weibull type.}}
    \label{figA3}
\end{figure}